# Amplified spontaneous emission at 5.23 μm in two-photon excited Rb vapour

A.M. Akulshin[1], N. Rahaman[1], S.A. Suslov[2] and R.J. McLean[1]

[1]Centre for Quantum and Optical Science, Swinburne University of Technology, PO Box 218, Melbourne 3122, Australia

[2]Department of Mathematics, Faculty of Science, Engineering and Technology, Swinburne University of Technology, Melbourne 3122, Australia

**Abstract.** Population inversion on the $5D_{5/2}$-$6P_{3/2}$ transition in Rb atoms produced by cw laser excitation at different wavelengths has been analysed by comparing the generated mid-IR radiation at 5.23 μm originating from amplified spontaneous emission (ASE) and isotropic blue fluorescence at 420 nm. Using a novel method for detection of two photon excitation via ASE, we have observed directional co- and counter-propagating emission at 5.23 μm. Evidence of a threshold-like characteristic is found in the ASE dependences on laser detuning and Rb number density. We find that the power dependences of the backward- and forward-directed emission can be very similar and demonstrate a pronounced saturation characteristic, however their spectral dependences are not identical. The presented observations could be useful for enhancing efficiency of frequency mixing processes and new field generation in atomic media.

## 1. Introduction

The interest in frequency conversion of resonant light and laser-like new field generation in atomic media specially prepared by resonant laser light is driven by a number of possible applications [1, 2, 3, 4]. The transfer of population to excited levels, particularly to create population inversion on strong transitions in the optical domain, is an essential part of the nonlinear process of frequency up- and down-conversion [5, 6, 7].

For example, two-photon excitation of Rb atoms from the ground state to the $5D_{5/2}$ level using laser fields at different wavelengths, as shown in Figure 1, produces population inversion on the three cascade one-photon transitions $5D_{5/2} \rightarrow 6P_{3/2}$, $6P_{3/2} \rightarrow 6S_{1/2}$ and $6S_{1/2} \rightarrow 5P_{1/2}$ since in each case the upper energy level has a longer lifetime than the lower level. If both the two-photon excitation rate and the atom number density $N$ are sufficiently high to produce more than one stimulated photon per spontaneous photon, then amplified spontaneous emission (ASE) might occur [8, 9]. In the highly-elongated interaction region defined by the applied laser beams this leads to directional stimulated emission at 1.32 μm and 5.23 μm [10, 11]. Intensities of these internally generated fields can be high enough to establish coherence between different states and subsequent parametric mixing with the applied laser radiation that produces coherent emission at 420 nm and 1.37 μm [5-7, 11] in directions allowed by the phase-matching condition.

Obviously, such process resulting in directional ASE could occur in other atomic species under appropriate laser excitation. Probably the first two-photon laser-induced ASE in pulsed and cw excited mercury vapours were observed in Refs. 12 and 13, respectively.

In this paper, we report on an experimental investigation of the mid-IR radiation at 5.23 μm generated due to the ASE process in two-photon excited Rb vapour. Spatial and spectral properties of ASE are defined to a large extent by the geometry of the region that contains population-inverted atoms [14]. While the intensity of isotropic resonant fluorescence is a good local indicator of the number of excited Rb atoms, the directional radiation originating from ASE provides complementary information about excited atoms integrated over the light-atom interaction region. ASE could also provide a signal with a higher signal-to-noise ratio compared to isotropic fluorescence when population inversion is prepared in a pencil-shaped interaction region within dilute vapours.

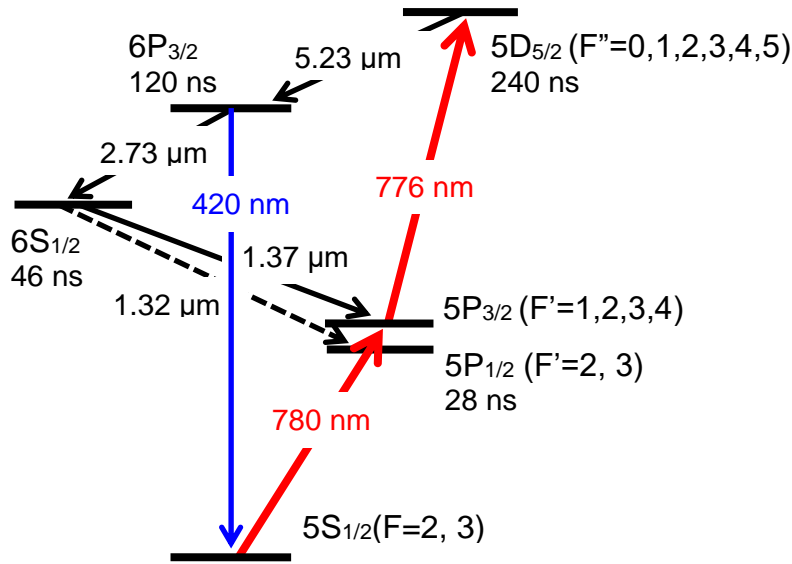

**Figure 1.** $^{85}$Rb atom energy levels involved in cascade population inversion.

Previous studies of this atomic system have relied on the detection of visible and near-IR radiation, from which the nature of the atomic processes is inferred. This is largely because of the difficulty of detecting the mid-IR radiation, which is strongly absorbed in conventional Rb cells.

We note that similar energy level configurations in Rb atoms have been used extensively for studying ladder-type electromagnetically induced transparency [15, 16], nonlinear properties of four-wave mixing and new field generation [17, 18], atomic coherence effects [19, 20] and transfer of orbital angular momentum [3, 21], as well as for imaging of ultracold atoms [22].

## 2. Two-photon excitation: General considerations

Doppler-free two-photon excitation [23] in alkali atoms is usually detected by observing isotropic fluorescence emitted during cascade decay. Typically, the collected fluorescence and applied laser light are spectrally separated, dramatically simplifying the detection procedure. Other popular methods are based on detecting absorption or polarization rotation of the applied laser light transmitted through an atomic sample [24, 25, 26]. Detection of new optical fields generated by ASE is less common [27].

There are two distinct ways of exciting Rb atoms to the $5D_{5/2}$ level using low-intensity cw resonant laser light at 780 and 776 nm.

- <u>Two-photon nearly velocity-insensitive excitation</u>

The excitation from the ground state to the $5D_{5/2}$ level can occur by absorbing two photons simultaneously when their sum frequency is resonant with the two-photon transitions: $\nu_{780}+\nu_{776}= \nu_{FF''}$, where $\nu_{FF''}$ are the frequencies of the transitions between the hyperfine sublevels F and F" of the $5S_{1/2}$ and $5D_{5/2}$ levels, respectively.

In the case of *counter-propagating* beams at 780 and 776 nm, almost the entire velocity distribution of Rb atoms interacts simultaneously with the applied laser light at the two-photon resonance, as the Doppler shift for moving atoms is almost compensated for the laser beams of similar wavelength:

$(v_{780} + v_{776}) + (k_{780} - k_{776})V/2\pi \approx v_{FF''}$, as $k_{780} \approx k_{776}$,

where $k_{780}$ and $k_{776}$ are the wave numbers at the appropriate wavelengths and $V$ is the velocity component of an atom in the direction of the 780 nm laser beam.

The two-photon excitation by counter-propagating beams is strongly enhanced when both optical frequencies are individually resonant to the one-photon transitions [28, 29].

- <u>Velocity-selective excitation</u>

Rb atoms can alternatively be excited to the $5D_{5/2}$ level within a single velocity group by *co-propagating* laser beams at 780 and 776 nm having approximately equal frequency detuning $\delta = (v_{780} - v_{FF'}) \approx (v_{776} - v_{F'F''}) < \Delta v_D$ from the corresponding one-photon transitions, where $\Delta v_D$ is the Doppler width. Atoms having $V \approx 2\pi \delta/k_{776} \approx 2\pi \delta/k_{780}$ velocity projection on the laser beam direction are at the two-photon resonance with the applied laser light despite the detuned sum frequency

$(v_{780} + v_{776}) - v_{FF''} = (k_{780} + k_{776})V/2\pi \approx 2\delta$.

In this velocity group Rb atoms could be excited to the $5D_{5/2}$ level by both incoherent stepwise and coherent two-photon processes. Numerical modelling of the excitation of Cs atoms in the corresponding energy level system showed that the contribution of the two-photon process prevails at higher laser intensity [24]. Velocity-selective and velocity-insensitive excitation produced by co- and counter-propagating radiation at 780 and 776 nm can be distinguished for the detuned radiation ($\delta \neq 0$), as the resonant condition for two processes are separated by $2\delta$.

In Rb vapours, narrow-linewidth fixed-frequency laser light at 780 nm that is tuned to the inhomogeneously broadened $^{85}$Rb-D2 absorption line interacts with atoms having certain velocity projections on the laser beam direction. Their velocity projections $V_{FF'}$ are determined by the frequency detuning of the laser $v_{780}$ from the corresponding resonant frequencies $v_{FF'}$ of the $5S_{1/2}$(F=2; 3) → $5P_{3/2}$(F'=2; 3; 4) hyperfine manifold:

$V_{FF'} = 2\pi(v_{780} - v_{FF'})/k_{780}$.

Under our experimental conditions when the 780 nm laser is tuned to the $5S_{1/2}$(F=3) → $5P_{3/2}$ manifold, the number of atoms in the $5S_{1/2}$(F=3) level in the $V_{32}$ and $V_{33}$ velocity groups is limited by optical pumping to the $5S_{1/2}$(F=2) level. Thus, Rb atom excitation to the intermediate $5P_{3/2}$ level predominantly occurs via the $5S_{1/2}$(F=3) → $5P_{3/2}$(F'=4) cycling transition, as was shown in Ref. 25 and the insignificant numbers of excited atoms in the $V_{32}$ and $V_{33}$ velocity groups are ignored in the following consideration.

## 3. Experimental setup

The experimental arrangement is similar to that used previously for studying two-photon spectroscopy [30] and directional polychromatic light generation [11] in Rb vapours.

Rb atoms are excited from the ground state to the $5D_{5/2}$ level by the optical fields from two extended-cavity diode lasers at 780 and 776 nm. The lasers are tuned close to the strong $5S_{1/2}$(F=3)→$5P_{3/2}$ and $5P_{3/2}$→$5D_{5/2}$ transitions in $^{85}$Rb (Figure 1). An additional laser tuned to the Rb-D1 line provides resonant radiation for ground-state repumping.

A gas of a natural mixture of Rb isotopes is contained in a 5-cm long heated glass cell with 1-mm thick sapphire windows, which are partially transparent at 5.23 μm. The temperature of the cell is set within the range 45-85°C, so that the atomic density $N$ of saturated Rb vapour could be varied between $1\times10^{11}$ cm$^{-3}$ and $12\times10^{11}$ cm$^{-3}$, where collisional broadening is negligible [31].

Rb atoms contained in the cell are simultaneously excited by a bichromatic beam that consists of carefully overlapped radiation at 780 and 776 nm and an almost counter-propagating laser beam at 776 nm (Figure 2). The laser power of all beams is controlled with wave plates and polarization beam splitters. The diameters of the beams are approximately 1.4 mm. The maximum laser power of all beams before entering the cell does not exceed 3 mW, so that the Rabi frequency $\Omega$ of the applied laser fields are $\gamma < \Omega < \Delta v_D$, where $\gamma$ is the natural linewidth of the one-photon transitions.

A room-temperature photovoltaic detector based on a variable-gap HgCdTe semiconductor with a 1x1 mm$^2$ light-sensitive area allows the 5.23 μm radiation to be detected using a lock-in amplifier. The isotropic blue fluorescence at 420 nm is detected at right angles to the direction of the applied laser beams by a photomultiplier tube (PMT). The blue fluorescence is separated from the scattered laser radiation and near-infrared fluorescence by colour filters.

The cell orientation is such that the reflection of applied laser light from the cell windows does not overlap with the incident beams. We find that birefringence of the sapphire cell windows results in minor polarization distortions of the transmitted laser light.

In our experiment, the ASE and blue fluorescence are recorded as a function of the optical frequency of either of the lasers while the frequency of the other laser remains fixed.

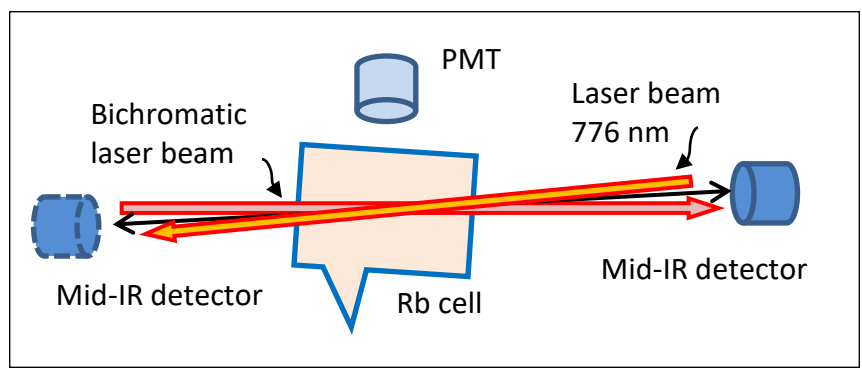

**Figure 2.** Simplified optical scheme of the experiment.

Polarization spectroscopy provides a convenient signal for modulation-free laser frequency locking not only to the strong transitions within the D-lines of alkali atoms, but also to optical transitions between excited levels [32, 26]. As was shown in Ref. 33, polarization resonances, which occur when the sum frequency of the 780 and 776 nm laser fields ($v_{780} + v_{776}$) is equal to the frequencies of the two-photon $5S_{1/2}(F) \to 5D_{5/2}(F")$ transitions ($v_{780} + v_{776} = v_{FF"}$), can be used as a reference for sum-frequency stabilization using a single servo system. However, sum frequency stabilization alone is not sufficient to ensure intensity-stable mid-IR directional emission as the excitation efficiency depends on frequency detuning from the one-photon transitions. Thus, the optical frequency of the 780 nm laser is independently modulation-free stabilized using the Doppler-free polarization resonance on the $5S_{1/2}(F=3) \to 5P_{3/2}(F'=4)$ transition or the nearest cross over resonance ensuring that $v_{780} = v_{34}$ or is red detuned $v_{780} \approx v_{34}$ - 60 MHz, respectively. Laser frequency stabilization results in reduced low-frequency variations of isotropic blue fluorescence and ASE at 5.23 μm. The measured relative intensity instability of the isotropic fluorescence at 420 nm within a 60 s time interval is less than 0.5%.

## 4. Experimental observations

Figure 3a shows typical spectral dependences of the forward-directed mid-IR radiation at 5.23 μm and isotropic blue fluorescence at 420 nm produced with the fixed frequency 780 nm laser detuned from the $5S_{1/2}(F=3) \to 5P_{3/2}(F'=4)$ transition, $v_{780} - v_{34} < \Delta v_D$. Both profiles, recorded simultaneously, reveal a structure that consists of two peaks separated by $2\delta$. The broader, lower peaks are the result of velocity-selective excitation produced by the co-propagating fields at 780 and 776 nm, while the higher and narrower peaks are due to two-photon excitation produced by the combination of the 780 nm component of the bichromatic beam and the counter-propagating radiation at 776 nm. In addition, both the mid-IR resonances originating from ASE on the population-inverted transition are noticeably narrower than the fluorescence peaks. We attribute this fact to a threshold-like gain effect in the population-inverted medium.

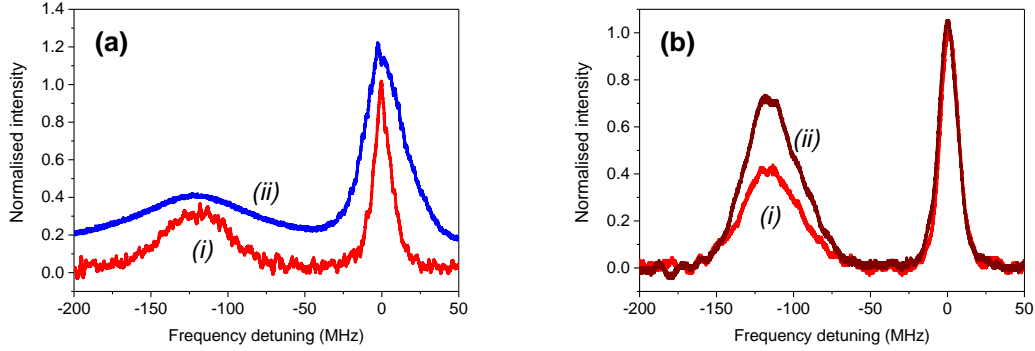

**Figure 3**. (a) Forward-directed ASE at 5.23 μm (*i*) and (*ii*) isotropic blue fluorescence at 420 nm (offset vertically by 0.2 divisions for clarity) as functions of the frequency detuning of the 776 nm laser from the two-photon resonance ($v_{780} + v_{776} = v_{35}$), while the fixed frequency 780 nm laser is approximately 60 MHz red-detuned from the $5S_{1/2}(F=3) \to 5P_{3/2}(F'=4)$ transition.
(b) Forward-directed ASE at 5.23 μm (i) without and (ii) with an additional repumping co-propagating beam ($I \approx 2$ mW/cm$^2$) (not shown in Fig. 2) that is tuned to the $5S_{1/2}(F=2) \to 5P_{1/2}(F'=3)$ transition at 795 nm, as a function of the 776 nm laser detuning from the two-photon resonance on the $5S_{1/2}(F=3) \to 5D_{5/2}(F''=5)$ transition. The 780 nm laser is locked approximately 60 MHz red-detuned from the $5S_{1/2}(F=3) \to 5P_{3/2}(F'=4)$ transition.

To emphasise the difference between the velocity-selective and velocity-insensitive excitations, we apply a co-propagating repumping laser beam at 795 nm (not shown in Fig. 2) that overlaps the bichromatic beam. A similar method was used for controlling nonlinear mixing processes and coherent blue light generation in alkali vapours [33]. If this laser light is sufficiently intense ($I > hv/\sigma_0 T$, where $\sigma_0$ is the absorption cross section of the transition in the $5S_{1/2} \to 5P_{1/2}$ manifold and $T$ is the interaction time determined by the beam diameter and the most probable velocity of Rb atoms [34]), then the fixed frequency laser field $v_{795}$ tuned to the vicinity of the open $5S_{1/2}(F=2) \to 5P_{1/2}(F'=2, 3)$ transitions produces efficient population repumping from the lower F=2 to the upper F=3 ground-state level in two velocity groups

$V_{23} = 2\pi(v_{795} - v_{23})/k_{795}$ and $V_{22} = 2\pi(v_{795} - v_{22})/k_{795}$,

where $V_{23}$ and $V_{22}$ are the velocity projections onto the direction of the laser beam, while $v_{795}$ and $k_{795}$ are the frequency and the wave number of the 795 nm laser radiation, respectively.

The number of atoms excited to the $5D_{5/2}$ level is significantly increased if the 795 nm laser is tuned to interact with the same velocity group as the 780 nm laser ($V_{23}=2\pi(v_{780} - v_{34})/k_{780}$ and $V_{22}=2\pi(v_{780} - v_{34})/k_{780}$). This in turn enhances both the velocity-selective and velocity-insensitive ASE peaks, as Figure 3b illustrates. However, the enhancements are different for the two methods of excitation. The velocity-selective peak is increased by approximately 70%. The velocity-insensitive or unmixed two-photon ASE peak, which is proportional to the total number of atoms regardless of their velocity projection, is enhanced by approximately 3%, as the number of pumped atoms is a small fraction of the total population of the F=3 level, approximately $\Gamma/\Delta v_D$, where $\Gamma$ is the homogeneous linewidth. In our experiment $\Gamma/\Delta v_D \approx 0.05$.

If the 780 nm and 795 nm laser fields interact with different velocity groups within the Doppler profile, the velocity selective ASE peak is unaffected by the presence of the D1-line laser, while the enhancement of the velocity-insensitive ASE peak depends weakly on the 795 nm laser detuning, as the total population in the F=3 level is always increased. The tuning of the repumping laser is not critical for the velocity-insensitive ASE, but it is for the velocity-selective mixed ASE peak originating from both stepwise and two-photon excitation processes. Although in general the enhancement is not exactly proportional to the increase in the number of resonant Rb atoms because of the threshold-like behaviour of the ASE, the ground-state hyperfine optical pumping observation does clearly demonstrate the difference between the velocity selective and velocity insensitive excitation to the $5D_{5/2}$ level using laser light at 780 and 776 nm.

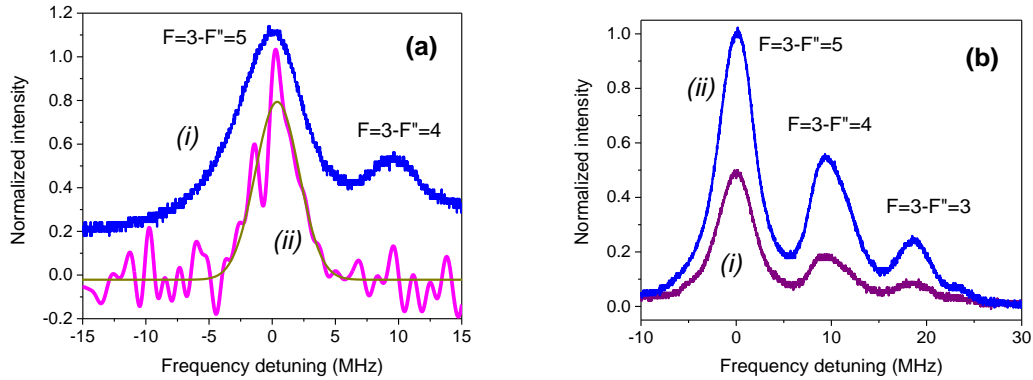

**Figure 4.** (a) Spectral dependence of (i) the blue fluorescence of the two-photon excited $^{85}$Rb atoms and (ii) the forward-directed ASE at 5.23 μm at low laser intensity (I < 1 mW/cm$^2$) and fitted to a Gaussian profile. The 780 nm laser is locked approximately 60 MHz red-detuned from the $5S_{1/2}(F=3) \rightarrow 5P_{3/2}(F'=4)$ transition. (b) Spectral dependence of (i) backward-directed and (ii) transverse blue fluorescence on the $5S_{1/2}(F=3) \rightarrow 5D_{5/2}(F")$ manifold.

It is worth noting that the velocity-selective two-wavelength excitation of Cs atoms with the equivalent energy level configuration ($6S_{1/2} \rightarrow 6P_{3/2} \rightarrow 6D_{5/2}$) was quantitatively analysed in Ref. 35. It was shown that only approximately 1% of the ground state population could reach the $6D_{5/2}$ using cw narrow-linewidth lasers.

In the spectral dependences presented in Figure 3a, the closely-spaced hyperfine structure of the $^{85}$Rb $5D_{5/2}$ level is not resolved, mainly due to power broadening. At lower laser power the two-photon peak of blue fluorescence reveals a partly resolved structure that corresponds to the excitation of $^{85}$Rb atoms from the $5S_{1/2}(F = 3)$ to $5D_{5/2}(F")$ levels allowed by the two-photon selection rule $\Delta F = |F - F"| \leq 2$ for the case of two photons with different frequencies [36], as shown in Figures 4b. However, the forward-directed ASE occurs only in the vicinity of the strongest two-photon $(F = 3) \rightarrow (F" = 5)$ transition (Figure 4a).

The narrowest observed full width at half maximum (FWHM) of the two-photon ASE resonance at such low intensity of the applied laser fields is approximately 4 MHz. The ultimate linewidth of the two-photon resonances produced by the monochromatic excitation is defined by a sum of the initial and final state decay rate $\gamma_S + \gamma_D$, where $\gamma_D \approx 2\pi \times 0.66$ MHz and $\gamma_S$ is determined by the time of flight of atoms across the laser beams. In the present case, these resonances are additionally broadened due to a mismatch in Doppler shifts $(v_{776}/v_{780} - 1)\Delta v_D \approx 2.5$ MHz arising from the frequency difference of the two optical fields.

We note that spectral dependences of blue fluorescence detected in the transverse and counter-propagating directions are practically identical (Figure 4b), while in the co-propagating direction the emission at 420 nm might occur in the form of intense spatially and temporally coherent light generated by parametric four-wave mixing [6, 7]. The similar spectral profiles of backward-directed and transverse blue fluorescence reflect the spontaneous emission origin. The phase matching condition is not satisfied in these directions. Also, the backward blue emission, unlike the mid-IR radiation, is not amplified after propagation through the medium because the $6P_{3/2} \rightarrow 5S_{1/2}(F=3)$ transition is not population inverted.

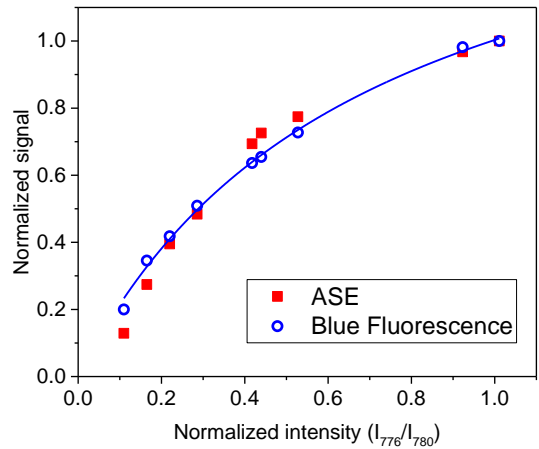

**Figure 5.** Normalized ASE at 5.23 μm and isotropic blue fluorescence and recorded simultaneously as a function of the 776 nm beam normalized power. The mid-IR emission is detected in the direction of the 780 nm beam. The experimental points are fitted to the $I/(1+I/I_{SAT})$ dependence. Optical frequency of the 780 nm laser is locked approximately 60 MHz red-detuned from the $5S_{1/2}(F=3) \rightarrow 5P_{3/2}(F'=4)$ transition, while the 776 nm laser is locked to the two-photon resonance on the $5S_{1/2}(F=3) \rightarrow 5D_{5/2}(F''=5)$ transition.

Figure 5 shows the intensity of the forward-directed emission at 5.23 μm and isotropic blue fluorescence as functions of the laser power at 776 nm. It is also worth noting that two-photon excitation and subsequent ASE at 5.23 μm could occur even with a large intensity imbalance between the counter-propagating beams at 780 and 776 nm. The dependences are similar and demonstrate a pronounced saturation characteristic. The saturation parameters $I_{SAT}$ evaluated via fitting the experimental data to the $I/(1+I/I_{SAT})$ dependence are similar, 63±4 and 67±13 mW/cm$^2$ for the isotropic blue fluorescence and mid-IR emission, respectively. A detailed study of power broadening of the ASE resonances and the dynamic Stark effect is the subject of a following paper.

The presented power dependence of the ASE does not reveal a sharp change in the rate of increase that typically occurs when the directional radiation appears from the isotropic spontaneous emission [8, 9], indicating the ASE threshold. The sensitivity of our detection system is not high enough to observe the isotropic fluorescence at 5.23 μm and the onset of the directional mid-IR radiation, i.e. the ASE threshold. The low collection efficiency results from the lack of a suitable lens for the 5.23 μm radiation and the small solid angle of the mid-IR detector.

Evidence of a threshold-like characteristic of ASE is found in the ASE dependence on 780 nm laser detuning and Rb number density. In Figure 6 we compare the variations of the isotropic blue fluorescence and the mid-IR emission detected in the direction of the 780 nm beam as a function of blue frequency detuning of the 780 nm laser from the $5S_{1/2}(F=3) \rightarrow 5P_{3/2}(F'=4)$ transition. High-frequency detuning is chosen to make contributions to the two-photon excitation from the weaker $5S_{1/2}(F=3) \rightarrow 5P_{3/2}(F'=2; 3)$ transitions negligible. The dependences are taken simultaneously at an atom number density $N \approx 5\times10^{11}$ $cm^{-3}$. Figure 6 shows that the mid-IR emission decays much faster with $\delta$; at $\delta > 400$ MHz the ASE signal is below the detection limit, while the blue fluorescence is still quite strong.

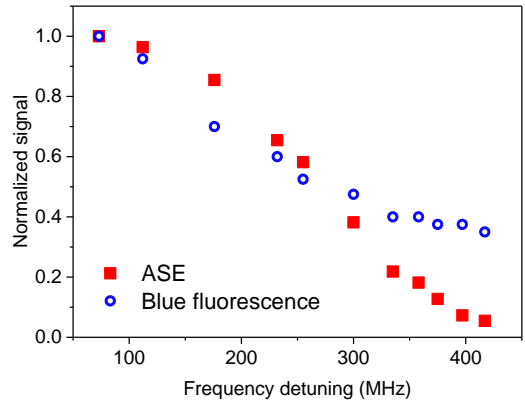

**Figure 6.** ASE emission originated from velocity-insensitive two-photon excitation and isotropic blue fluorescence detected simultaneously as a function of high frequency detuning of the 780 nm laser from the $5S_{1/2}(F=3) \to 5P_{3/2}(F'=4)$ transition, $\delta = v_{780} - v_{34}$, but with the sum of laser frequencies ($v_{780} + v_{776}$) matching the frequency of the $5S_{1/2}(F=3) \to 5D_{5/2}(F''=5)$ two-photon transition. ASE at 5.23 μm is detected in the direction of the 780 nm beam.

Figure 7 demonstrates the Rb number density dependence of the two-photon ASE and isotropic blue fluorescence. Linear fits to the experimental data presented in the log-log plot reveal that the isotropic blue fluorescence is proportional to the Rb number atomic density $N$ ($\sim N^\beta$, where $\beta = 0.95 \pm 0.04$), while the ASE demonstrates much steeper rise ($\beta = 2.3 \pm 0.2$) at lower Rb number density ($N < 1.1 \times 10^{11}$ $cm^{-3}$) followed by the region of approximatly linear growth.

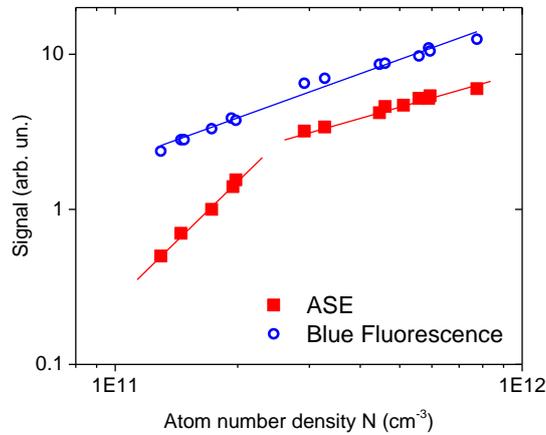

**Figure 7.** Two-photon ASE and isotropic blue fluorescence as a function of Rb number density $N$. Mid-IR light is detected in the direction of the 780 nm beam. The 780 nm laser is approximately 60 MHz red-detuned from the $5S_{1/2}(F=3) \to 5P_{3/2}(F'=4)$ transition, while the 776 nm laser is locked to the two-photon resonance on the $5S_{1/2}(F=3) \to 5D_{5/2}(F''=5)$ transition.

The observation of blue fluorescence without detecting mid-IR emission at 5.23 μm at some experimental conditions, such as large frequency detuning $\delta$ (Fig. 6) or low atom number density $N$ (Fig. 7) might seem counterintuitive, as the population transfer from the $5D_{5/2}$ to $6P_{3/2}$ levels implies mid-IR photons are emitted. This could be explained simply by different detection thresholds for the mid-IR and blue fluorescence. The intensity of isotropic radiation at 5.23 μm is below our detection limit, but it is present, and results in blue fluorescence that is detected with much higher efficiency. Indeed, the detected blue fluorescence might result entirely from spontaneous process on the $5D_{5/2} \to 6P_{3/2}$ transition, while the detection sensitivity is not sufficient to detect the isotropic fluorescence at 5.23 μm.

We find that under a range of experimental conditions the backward- and forward-directed ASE generated in Rb vapours velocity-selective excited entirely by nearly counter-propagating fields at 780 and 776 nm exhibit similar power dependences. The plots of spectral dependences of ASE in the backward and forward directions are shown in Figure 8. Despite very similar intensities observed in both directions, their spectral profiles are not identical. The mid-IR emission in the direction of the 776 nm beam is shifted toward lower frequencies by approximately half the width of the spectral profile.

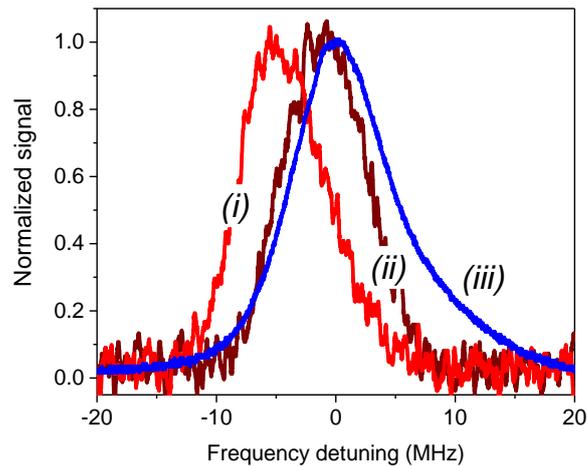

**Figure 8**. ASE and isotropic blue fluorescence generated in Rb vapours excited by counter-propagating laser beams (curves i and ii) as functions of the frequency detuning of the 776 nm laser from the two-photon resonance ($v_{780} + v_{776} = v_{35}$), while the 780 nm laser is locked to the $5S_{1/2}(F=3) \rightarrow 5P_{3/2}(F'=4)$ transition. Curves (i) and (ii) represent the radiation at 5.23 μm detected in the direction of the 776 nm and 780 nm beams, respectively, while curve (iii) shows the variation of isotropic blue fluorescence.

We emphasise that the velocity-selective population inversion on the $5D_{5/2} \rightarrow 5P_{3/2}(F=3)$ transition results in very distinctive power dependences in the co- and counter-propagating directions as was also shown in [11]. A similar intensity imbalance was observed in three-photon pumped argon-air mixtures and attributed to the asymmetry of the gain in the excitation region [37]. It has been suggested that the backward- and forward-propagating emissions compete for gain [38].

The distinctive spectral and intensity features of the laser-like backward and forward emission in Rb vapour excited simultaneously by stepwise and two-photon process or entirely via the resonant two-photon process merit more detailed investigation because of their possible applicability to remote sensing.

## 5. Conclusions

Population inversion on the $5D_{5/2} \rightarrow 6P_{3/2}$ transition in Rb atoms produced by two-photon excitation at different wavelengths has been analysed by comparing mid-IR emission at 5.23 μm and isotropic blue fluorescence at 420 nm. The intensity of isotropic resonant fluorescence is a good local indicator of the number of excited Rb atoms, while the directional radiation originating from ASE provides complementary information about excited atoms integrated over the light-atom interaction region.

It is shown that depending on the geometry of the applied radiation at 780 and 776 nm the excitation could be predominantly velocity-selective or velocity-insensitive, producing spectrally and spatially distinguishable directional radiation at 5.23 μm via the mechanism of ASE. Thus, a new way of detecting and analysing two-photon excitation in atomic vapours using ASE is suggested.

Specific features of velocity-insensitive and velocity-selective excitation have been emphasised using a ground-state population repumping approach.

Intensity dependences of directional mid-IR emission at 5.23 μm and isotropic blue fluorescence at 420 nm reveal a saturation characteristic. We find that the saturation intensities for both cases are similar.

Spectral profiles of ASE for both the velocity-selective and velocity-insensitive excitation are narrower than the corresponding resonances of isotropic blue fluorescence, which is attributed to the amplification effect. The smallest observed full width of the two-photon ASE resonance is approximately 4 MHz.

Analysing intensity and spectral dependences confirms that the directional mid-IR radiation is due to stimulated process on a population-inverted transition, while the isotropic blue fluorescence originates from spontaneous relaxation. Better understanding the specific properties of directional mid-IR emission is important for optimizing the efficiency of frequency mixing processes and new field generation in atomic media.

Finally, ASE-based detection might be more suitable for return signal remote sensing, as this emission occurs in both forward and backward directions. The mechanism of ASE described here could be easily extended to other atomic species allowing generation of coherent fields at exotic wavelengths.

**Acknowledgments**

The authors are grateful to Dmitry Budker for his interest, as well as for making the Rb cell with sapphire windows and the mid-IR detector available for our experiments. AA also warmly thanks Tatiana Tchernova for technical assistance.

**References**

[1] Hebert T, Wannemacher R, Lenth W and Macfarlane R M 1990 *Appl. Phys. Lett.* **57** 1727
[2] Radnaev A G, Dudin Y O, Zhao R, Jen H H, Jenkins S D, Kuzmich A and Kennedy T A B 2010 *Nature Phys.* **6** 894
[3] Walker G, Arnold A S and Franke–Arnold S 2012 *Phys. Rev. Lett.* **108** 243601
[4] Moran P J, Richards R M, Rice C A and Perram G P 2016 *Opt. Commun.* **374** 51
[5] Zibrov A S, Lukin M D, Hollberg L and Scully M O 2002 *Phys. Rev. A* **65** 051801
[6] Akulshin A M, McLean R J, Sidorov A I and Hannaford P, 2009 *Opt. Express* **17** 22861
[7] Vernier A, Franke-Arnold S, Riis E, and Arnold A S 2010 *Opt. Express* **18** 17026
[8] Peters G I and Allen L 1971 *J. Phys. A: Gen. Phys* **4** 238
[9] Casperson L W 1977 *J. Appl. Phys.* **48** 256
[10] Sell J F, Gearba M A, DePaola B D and Knize R J 2014 *Opt. Lett.* **39** 528
[11] Akulshin A, Budker D and McLean R 2014 *Opt. Lett.* **39** 845
[12] Omenetto N, Matveev O I, Resto W, Badini R, Smith B W, and Winefordner J D 1994 *Appl. Phys. B* **58** 303
[13] Kolbe D, Scheid M, Koglbauer A and Walz J 2010 *Opt. Lett.* **35** 162690
[14] Allen L and Peters G I 1972 *J. Phys. A: Gen. Phys.* **5** 546
[15] Sargsyan A, Sarkisyan D, Krohn U, Keaveney J and Adams C 2010 *Phys. Rev. A* **82** 045806
[16] Keaveney J, Sargsyan A, Sarkisyan D, Popoyan A and Adams C 2014 *J. Phys. B* **47** 075002
[17] Parniak M and Wasilewski W, 2015 *Phys. Rev. A* **91**, 023418
[18] Sulham C V, Pitz G A and Perram G P 2010 *Appl. Phys.* B **101** 57
[19] Lee H-g, Kim H, Lim J and Ahn J 2013 *Phys. Rev. A* **88** 053427
[20] Lee Y-S and Moon H S, 2016 *Opt. Express* **24** 10723
[21] Akulshin A M, Novikova I, Mikhailov E E, Suslov S A and McLean R J 2016 *Opt. Lett.* **41**, 1146
[22] Ohadi H, Himsworth M, Xuereb A and Freegarde T 2009 *Opt. Express* **17,** 23003
[23] Vasilenko L S, Chebotaev V P and Shishaev A V 1970 *JETP Lett.* **12** 113

24 Kargapol'tsev S V, Velichansky V L, Yarovitsky A V, Taichenachev A V and Yudin V I 2005 *Quant. Electron.* **35** 591
25 Hamid R, Cetintac M and Celik M 2003 *Opt. Commun.* **224** 247
26 Parniak M, Leszczynski A and Wasilewski W 2016 *Appl. Phys. Lett.* **108** 161103
27 DeBoo B, Kimball D K, Li C H and Budker D 2001 *JOSA B* **18** 639
28 Salomaa R and Stenholm S 1975 *J. Phys. B* **8** 1795
29 Salomaa R and Stenholm S 1976 *J. Phys. B* **9** 1221
30 Akulshin A M, Hall B V, Ivannikov V, Orel A A and Sidorov A I 2011 *J. Phys. B* **44** 215401
31 Akulshin A M, Celikov A A, Sautenkov V A, Vartanian T A and Velichansky V L 1991 *Opt. Commun.* **85** 21
32 Pearman C P, Adams C S, Cox S G, Griffin P F, Smith D A and Hughes I G 2002 *J. Phys. B* **35** 5141
33 Akulshin A M, Orel A A and McLean R J 2012 *J. Phys. B* **45** 015401
34 Pappas P G, Burns M M, Hinshelwood D D, Feld M S and Murnick D E 1980 *Phys. Rev. A* **21** 1955
35 Horvatic V, Correll T L, Omenetto N, Vadla C and Winefordner J D 2006 *Spectrochim. Acta B* **61** 1260
36 Bonin K D and McIlrath T J 1984 *J. Opt. Soc. Am. B* **1** 52
37 Dogariu A and Miles R B 2016 *Opt. Express* **24** A544
38 Laurain A, Scheller M and Polynkin P 2014 *Phys. Rev. Lett.* **113** 253901